\documentclass[preprint]{aastex}
\usepackage{emulateapj5}
\usepackage{graphicx}
\usepackage{rotate}
\usepackage{amsmath}
\usepackage{amssymb}

\begin{document}

\title{Flux tubes as the origin of Net Circular Polarization 
in Sunspot Penumbrae}

\author{J.M.~Borrero}
\affil{High Altitude Observatory (NCAR), 3080 Center Green Dr. CG-1, Boulder, CO 80301, USA}
\email{borrero@ucar.edu}
\author{L.R.~Bellot Rubio}
\affil{Instituto de Astrof{\'\i}sica de Andaluc{\'\i}a (CSIC), Apdo. 2004, 18080 Granada, Spain}
\email{lbellot@iaa.es}
\and
\vspace{-0.5cm}
\author{D.A.N.~M\"uller}
\affil{European Space Agency c/o NASA Goddard Space 
Flight Center, Mailcode 671.1, Greenbelt, MD 20771, USA}
\email{dmueller@esa.nascom.nasa.gov}

\begin{abstract}
{We employ a 3-dimensional magnetohydrostatic model of
a horizontal flux tube, embedded in a magnetic surrounding
atmosphere, to successfully reproduce the azimuthal and
center-to-limb variations of the Net Circular Polarization
observed in sunspot penumbrae. This success is partly due to the
realistic modeling of the interaction between the flux tube
and the surrounding magnetic field.}
\end{abstract}

\keywords{Sun: sunspots -- Sun: magnetic fields -- Sun: polarimetry}

\shorttitle{NCP in penumbral flux tubes}
\shortauthors{Borrero et al.}
\maketitle

\def\nn{{\bf \nabla}}
\def\cro{\times}
\def\er{{\bf{\rm e_{\rm r}}}}
\def\et{{\bf{\rm e_{\rm \theta}}}}
\def\ex{{\bf{\rm e_{\rm x}}}}
\def\ey{{\bf{\rm e_{\rm y}}}}
\def\ez{{\bf{\rm e_{\rm z}}}}
\def\nx{\mathcal{N}}
\def\sx{\mathcal{S}}
\def\rx{\mathcal{R}}

\section{Introduction}

A possible scenario that explains the magnetic field configuration
of the sunspot penumbra is the so-called uncombed penumbral model (Solanki \& 
Montavon 1993). In this model a horizontal flux tube that harbors
the Evershed flow is embedded in a more vertical magnetic field.
A similar configuration has been obtained by Heinemann et al. (2007),
who carried out 3D MHD simulations of the penumbra and found
that penumbral filaments are produced by bubbles of weak and horinzontal 
magnetic field fully embedded in a stronger and more vertical one.
They also find field free gaps connected to the deeper convection zone
which appear as bright regions in the emergent intensity.

One of the most critical observations that any penumbral model must 
reproduce, is the Net Circular Polarization (S\'anchez Almeida \& Lites 1992).
Different realizations of the uncombed penumbra have successfully
explained these observations (Solanki \& Montavon 1993; Mart{\'\i}nez Pillet 2000;
Schlichenmaier et al. 2002; M\"uller et al. 2002, 2006). However, 
Spruit \& Scharmer (2006) have pointed out that the models of the uncombed penumbra 
published so far, do not consider the effects of the external field wrapping 
around the flux tube. The goal of this work
is to remove this point of inconsistency. As it will be shown, a 
uncombed model that consistently considers the perturbation in the
external field introduced by a cylindrical flux tube, is also able to
explain the Net Circular Polarization observed in the sunspot penumbra with 
surprising accuracy.

\section{Flux tube model and reference frame}

Our embedded flux tube model is adopted from Borrero (2007).
This model considers a cylindrical flux tube that carries the
Evershed flow. The flux tube is located perpendicularly
to the vertical $z$-axis and is embedded in an inclined
potential magnetic field.

This model prescribes the magnetic field configuration in the local 
reference frame (LRF): $\sx=\{\ex,\ey,\ez\}$; see Fig.~1),
being the flux tube's axis oriented along $\ex$.
Analytical expressions for $B_x(y,z)$, $B_y(y,z)$, and $B_z(y,z)$
are taken from Eqs.~(33)-(34) in Borrero (2007).
These equations take into account how the external field bends and 
surrounds the flux tube. Although the model is three-dimensional, variations along the 
$x$-axis are neglected. The model is described by 6 
parameters: $B_0$ and $\gamma_0$ (strength and inclination of the external magnetic field
far away from the flux tube), $R$ and $z_0$ (radius and central position of the flux tube), 
and finally $B_{xt0}$ and $v_{xt0}$ (component of the flux tube magnetic 
field along the flux tube's axis and magnitude of the Evershed flow).
Figure 1 (top panel) illustrates the magnetic field lines in the plane perpendicular to 
the tube's axis.

The velocity and magnetic field vectors enter the static
momentum equation. This yields the density $\rho(y,z)$,
and gas pressure $P_g(y,z)$, that  ensure force balance (see Borrero 2007 for 
details). Once gas pressure and density are known, the temperature $T(y,z)$ is
evaluated through the equation of state for ideal gases. A varying molecular 
weight is used to account for the partial ionization of the different atomic species.

Once all the relevant quantities are known in the LRF, $\sx$,
we project the magnetic field and velocity vectors on the observer's
reference frame (ORF), $\sx{''}$. This is accomplished by a
rotation of angle $\Psi$ along the vertical $z$-axis. This
rotation ensures that the resulting $x^{'}$-axis meets the line of symmetry of 
the sunspot. A second rotation of angle $\Theta$ 
along $\ey^{'}$ is finally needed to direct the resulting $z^{''}$
 along the observer's line-of-sight. Mathematically,

\begin{equation}
\sx^{''}=\{\ex^{''},\ey^{''},\ez^{''}\}=\rx_{y^{'}}(\Theta)
\times \rx_z(\Psi) \times \sx {\rm\;,}
\end{equation}\

\begin{center}
\includegraphics[width=8.5cm]{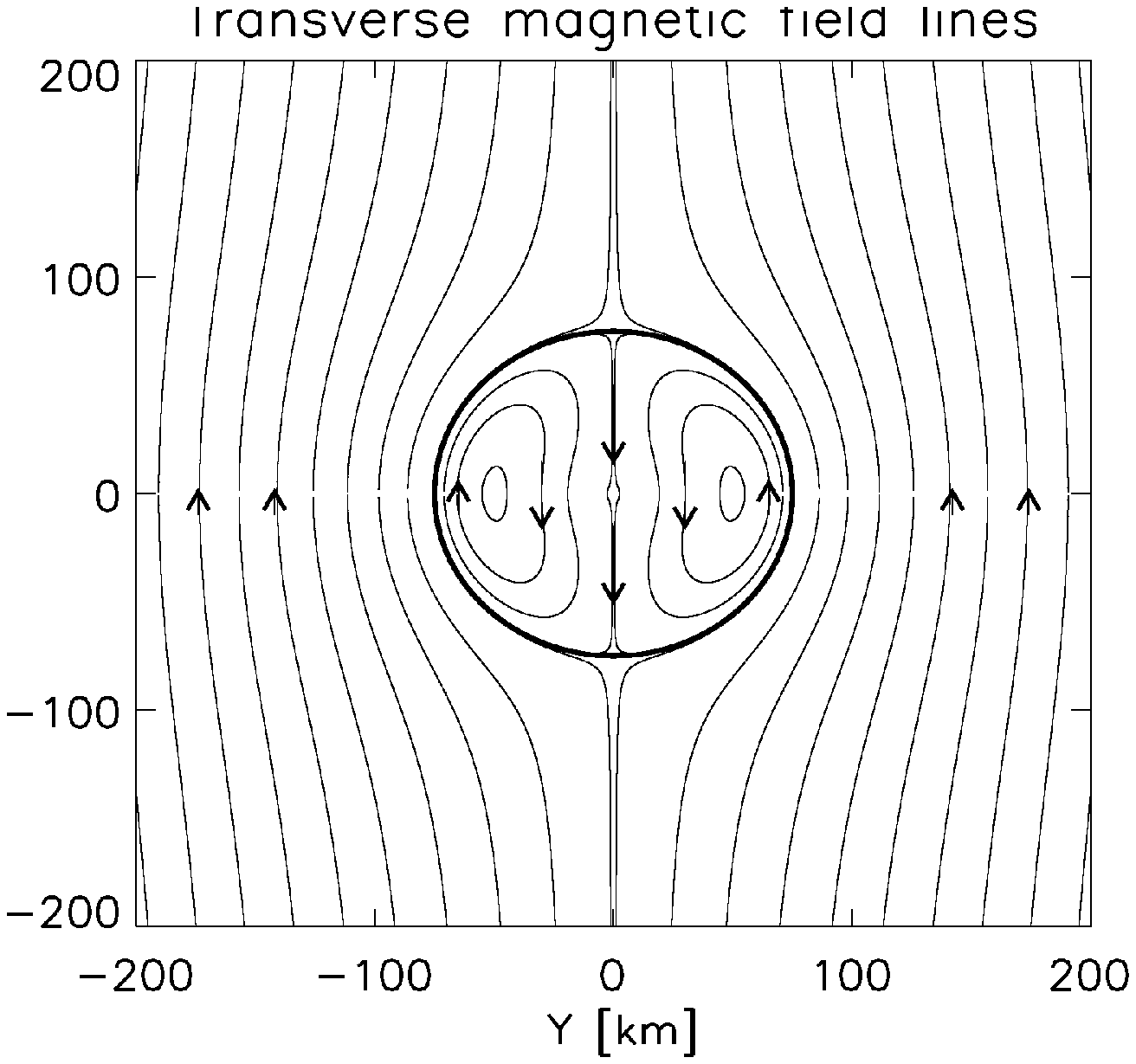}
\includegraphics[width=8.5cm]{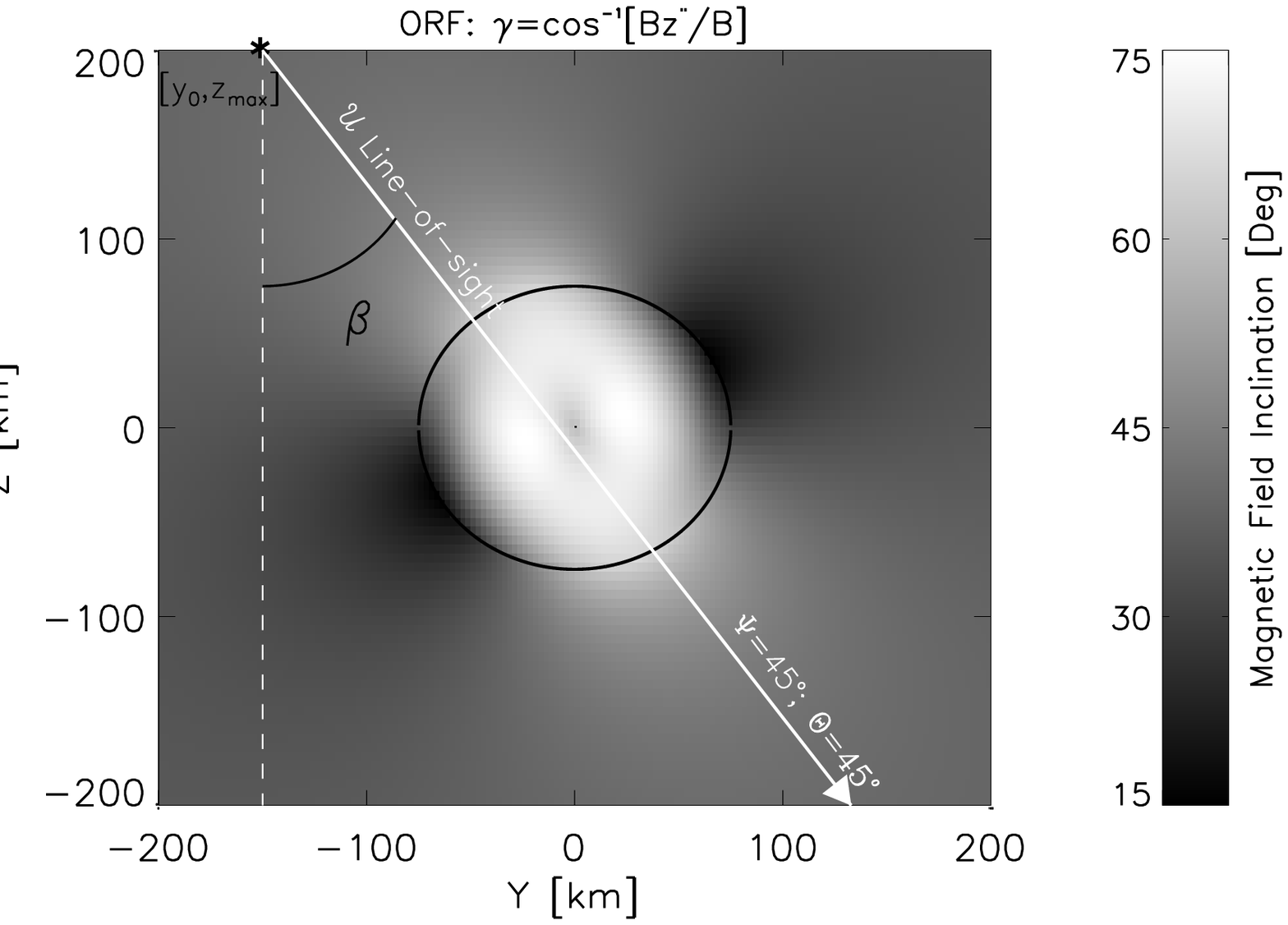}
\figcaption{{\it Top panel}: magnetic field lines tranversal to
the flux tube's axis. Note that inside the flux tube the magnetic field is 
mostly aligned its axis and therefore it is not shown here.
{\it Bottom panel}: inclination of the magnetic field in the observer's
reference frame (i.e.: inclination respect to the line-of-sight).
The white arrow represents a possible line-of-sight of an observer at
$\Theta=45^{\circ}$ looking at a penumbral flux tube located at $\Psi=45^{\circ}$
with respect to the line of symmetry in the center side. In this example, the intersection of the
ray-path with the uppermost point in the $yz$ plane occurs at ($y_0,z_{max}$)=($-150,200$)
[km]. This point is indicated by a black asterisk. The
angle between the vertical $z$-axis (white dashed line) and the projected line-of-sight is
$\beta \simeq 35^{\circ}$. Other ray-paths would be parallel to the one drawn but 
intersect at different $y_0$'s.}
\end{center}

\noindent where $\rx_{y^{'}}(\Theta)$ and $\rx_{z}(\Psi)$ are
the corresponding rotation matrices. In this manner, we can locate the flux tube 
at any azimuthal position $\Psi$ within the sunspot, and to
place the sunspot at any heliocentric angle, $\Theta$, on the solar disk. 
Thus, we can determine the line-of-sight velocity $v_{\rm los}=v_z^{''}$, 
magnetic field strength $B$ (calculated in any reference frame), 
inclination of the magnetic field with respect to the line-of-sight 
$\gamma=\cos^{-1} [B_z^{''}/B]$, and the angle of the magnetic field vector 
in the plane perpendicular to the line-of-sight $\phi = \tan^{-1} [B_x^{''}/B_y^{''}]$. 
An example of $\gamma(y,z)$ is presented in Fig.~1 (bottom panel).

Finally, our physical parameters are expressed as function of $(y,z)$, but
for our radiative transfer calculations we need to know
them along the line-of-sight. To this end, we project the
observer's line-of-sight onto the $yz$ plane: $\mathcal{U}=\sin\Psi\sin\Theta\ey-
\cos\Theta\ez$. With this vector we can construct the parametric equation
of the line-of-sight in the $yz$ plane:

\begin{equation}
(y,z) = (y_0,z_{max}) + \lambda \; (\sin\Psi\sin\Theta, -\cos\Theta) 
\end{equation}

\noindent where $(y_0,z_{max})$ is the uppermost
intersection point of the ray-path with the $yz$ plane.
The slope-intercept form of the line-of-sight is:

\begin{equation}
y = y_0+(z_{max}-z) \tan\Theta \sin \Psi
\end{equation}

Note that, either at disk center ($\Theta=0$) or along to the
line of symmetry of the sunspot ($\Psi=0,\pi/2$), the ray-paths
are given by $y=y_0$. At any other position, the ray-paths enter
the $yz$ plane, forming an angle $\beta = \tan^{-1}(\tan\Theta \sin \Psi)$
with the vertical direction. As expected, $\beta=\pm \Theta$
perpendicularly to the line of symmetry (\ $\Psi=\pi/2,3\pi/2$).
An example of the projection of the line-of-sight onto the
$yz$ plane is presented in Fig.~1.

After these geometrical considerations, we are now ready to 
solve the radiative transfer equation and obtain theoretical  
Stokes profiles from our embedded flux tube model.
We have employed three different numerical codes:
SIR (Ru{\'\i}z Cobo \& Del Toro Iniesta 1992),
DIAMAG (Grossmann-Doerth 1994) and SPINOR (Frutiger 2000)
and verified that the results are consistent among them.
We first calculate the emerging polarization profiles for 128 rays 
that enter into the $yz$ plane with different $y_0$'s (see Fig.~1).
We then compute the Net Circular Polarization (NCP) as the wavelength
integral of Stokes $V$. The total NCP is obtained as sum of the NCP 
produced by each individual ray-path (see example in Fig.~2; top panel).

\section{Observations}

Five sunspots at five different positions on the solar disk
were observed using the Tenerife Infrared Polarimeter (TIP; Mart{\'\i}nez Pillet et al. 1999) 
at the Vacuum Tower Telescope (Iza\~na Observatory, Spain), and the Advanced Stokes Polarimeter 
(ASP; Elmore et al. 1992) at the Dunn Solar Telescope (Sacramento Peak, USA). The full Stokes vector 
of \ion{Fe}{1} 15648.5 \AA\ (TIP) and \ion{Fe}{1} 6302.5 \AA\ (ASP) were recorded .
For each sunspot, radially averaged azimuthal variations of the NCP, $\nx(\Psi)$, are computed. 
Here $\Psi$ runs counter-clockwise, with $\Psi=0$ referring to the line of symmetry on 
the center side of the penumbra. The determination of the position of the line of symmetry 
from the observations is a difficult task and can only be done to an accuracy of about $\pm 10^{\circ}$.


\begin{center}
\tabcaption{Summary of observed sunspots. References:
{\bf a} (Cabrera Solana et al. 2006), {\bf b} (Borrero et al. 2006), 
{\bf c} (Bellot Rubio et al. 2004), {\bf d} (Borrero et al. 2005).}
\begin{tabular}{|ccccc|}
\tableline
AR & Date & $\Theta$ [Deg] & $\lambda$ [\AA] & Instrument \\
\tableline
10430 & 08Aug03 & 27 & \ion{Fe}{1} 15648 & TIP(a)\\
8545 & 21May99 & 38 & \ion{Fe}{1} 6302 & ASP(b)\\
8704 & 20Sep99 & 40 & \ion{Fe}{1} 15648 & TIP(c)\\
10425 & 03Aug03 & 50 & \ion{Fe}{1} 15648 & TIP(a)\\
8706 & 21Sep99 & 60 & \ion{Fe}{1} 15648 & TIP(d)\\  
\tableline
\end{tabular}
\end{center}

\begin{center}
\includegraphics[angle=0,width=7.5cm]{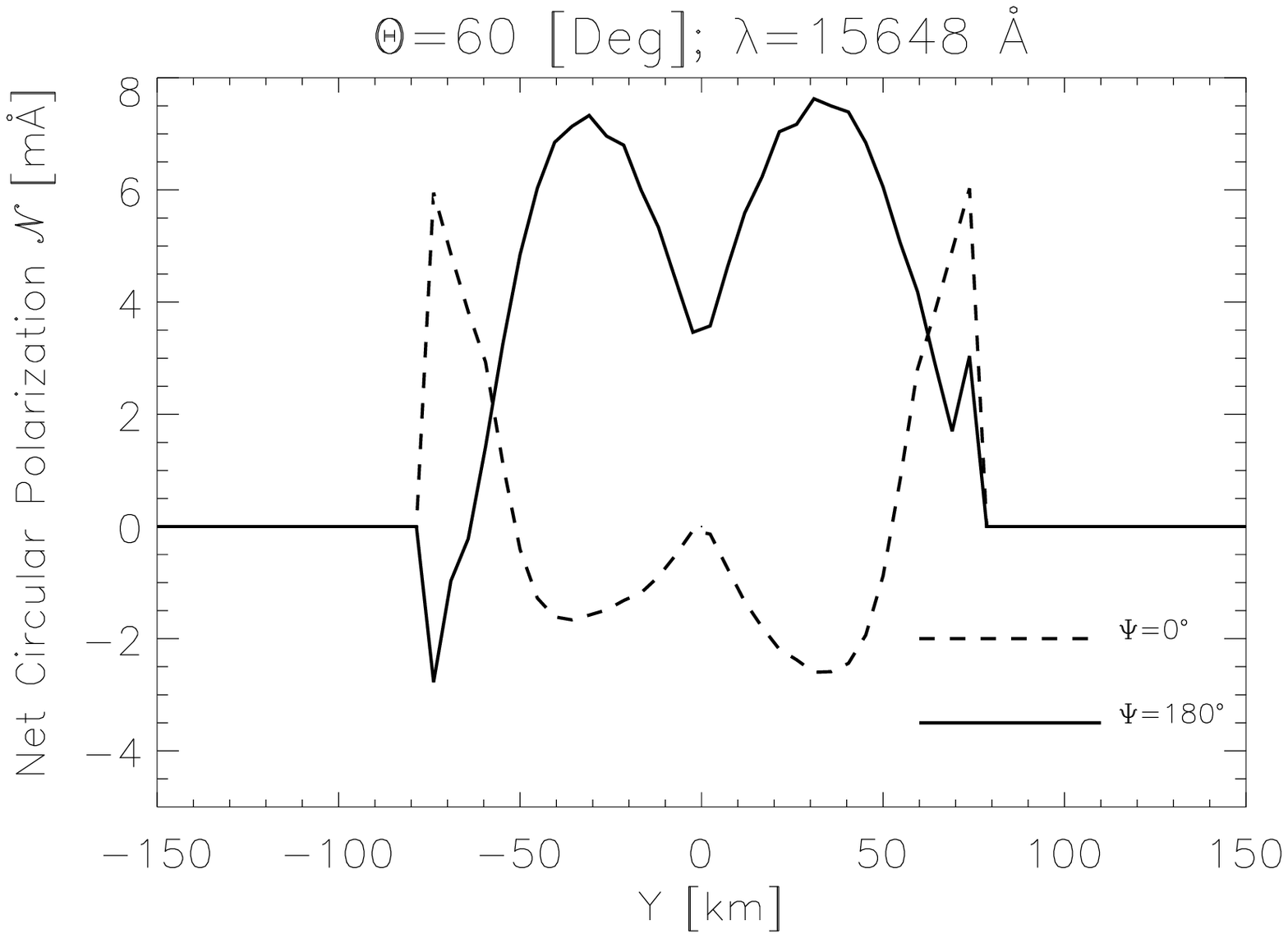}
\includegraphics[angle=0,width=7.5cm]{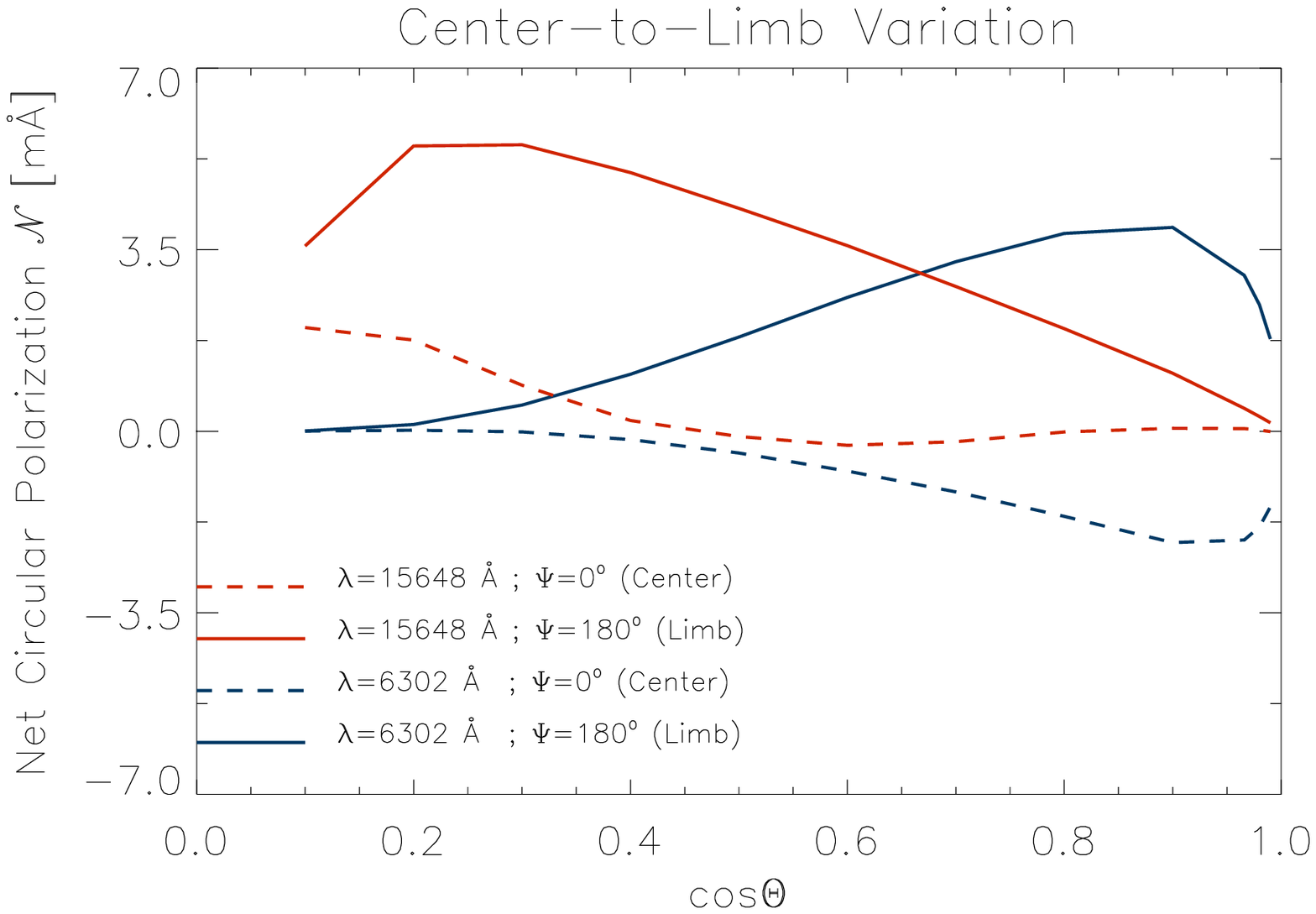}
\figcaption{{\it Top panel}: Net Circular Polarization
produced by each ray-path piercing the flux tube
at different $y$ positions. The flux tube is located 
at the line of symmetry on the center side 
(dashed, $\Psi=0$) and on the limb side (solid; $\Psi=\pi$).
{\it Bottom panel}: variation of the total NCP (integral
over $y$ in the top panel) as a function of the heliocentric position
of the sunspot: center-to-limb-variation.}
\end{center}

\section{Observed vs Theoretical NCP}
\subsection{Center-to-limb variation}

We have computed the NCP that emerges from our flux tube model for different heliocentric
angles $\Theta$. This is the so-called center-to-limb variation of the NCP. 
We have repeated this calculation for flux tubes located at the line of symmetry of the sunspot, 
on the center side ($\Psi=0$) and on the limb side ($\Psi=\pi$). The NCP
was evaluated for the same spectral lines of our observations (Sect.~3).
Results are presented in Fig.~2 (bottom panel). The parameters used for the flux tube model 
are: $B_0=B_{xt0}=1000$ Gauss, $\gamma_0=60^{\circ}$, $v_{xt0}=6$ km s$^{-1}$, $R=75$ Km,
 $z_0=0$ km. These values are consistent with result from spectropolarimetric
observations of the penumbral fine structure. In addition, we have used the hot 
umbral model by Collados et al. (1994) to represent
 the thermodynamic parameters of the atmosphere surrounding the flux tube.

Mart{\'\i}nez Pillet (2000) has presented the observed center-to-limb variation of 
the NCP in \ion{Fe}{1} 6302.5 \AA\ for a large number of sunspots.
Our predictions (Fig.~2; bottom panel; blue lines) agree very well with his findings.
Our theoretical $\nx(\Theta)$ curve for 6302.5 \AA\
does not cross zero at $\cos \Theta=[0.8,1]$, however. This is due to
our model simplifications: flux tube is always 
perpendicular to the vertical $z-$axis, and the external 
atmosphere does not harbor any flows.

For the near infrared neutral iron line at 15648.5 \AA\, we
are not aware of any similar work to that of Mart{\'\i}nez Pillet.
However, our theoretical curve is in very good agreement with
other theoretical predictions that use a simpler uncombed
scenario (see M\"uller et al. 2002, 2006). The reason is that,
along the line of symmetry ($\Psi=0,\pi$), ray-paths
are not inclined on the $yz$ plane regardless of
the heliocentric angle (Eq.~3). 

\subsection{Azimuthal variations}

In this section we compare the observed
$\nx(\Psi)$, for different sunspots
at different heliocentric angles $\Theta$,
with the theoretical predictions from
the embedded flux tube model. This is done
individually for each sunspot. Theoretical curves have 
been obtained using the same model parameters as in Sect.~4.1. 
Results are presented in Figure 3. It can be 
seen that our model is able to reproduce many features
of the observed azimuthal variations of the NCP.
This achievement is especially remarkable
if we consider that all we have done is changing
the spectral line and heliocentric angle (model parameters
were kept constant).

Particularly interesting is the fact that
we can also predict the existence
of secondary maxima/minima in the $\nx(\Psi)$ curves.  
This is clearly the case of Fe {\rm I} 15648.5 \AA~, where 
more simple uncombed models, predict only the existence of 
2 maxima and 2 minima at all heliocentric angles.
The improved agreement between observations and predictions
is to be ascribed to our more realistic modeling of the external 
magnetic field bending and wrapping around the horizontal flux tube.
This is a crucial ingredient for \ion{Fe}{1} 15648.5 \AA~, where 
gradients in azimuthal angle of the magnetic field play a major role 
(Landolfi \& Landi degl'Innocenti 1996; M\"uller et al. 2002).

An important detail is that, in our
model, the Evershed flow is channeled
along the horizontal flux tube. This means that the line
of sight velocity, $v_{los}=v_z^{''}=-v_{xt0} \cos \Psi \sin \Theta$, vanishes
always perpendicularly to the line of
symmetry ($\Psi=\pi/2,3\pi/2$). Thus, 
gradients in $v_{los}$ do not exist there, yielding
always zero Net Circular Polarization.
In agreement with M\"uller (2001), decreasing the magnitude of the Evershed
 flow decreases the amount of NCP roughly linearly. In the absence of
a complete parameter study, this effect alone
cannot be used to rule out weaker horizontal flows, however.

Borrero (2007) expressed concerns about flux tubes 
with circular cross sections having very smooth variations in $\gamma$. 
He pointed out that those variations might be too small to generate enough 
NCP (S\'anchez Almeida \& Lites 1992). Results presented in this work
prove those concerns to be unfounded.

\begin{center}
\includegraphics[angle=0,width=7.5cm]{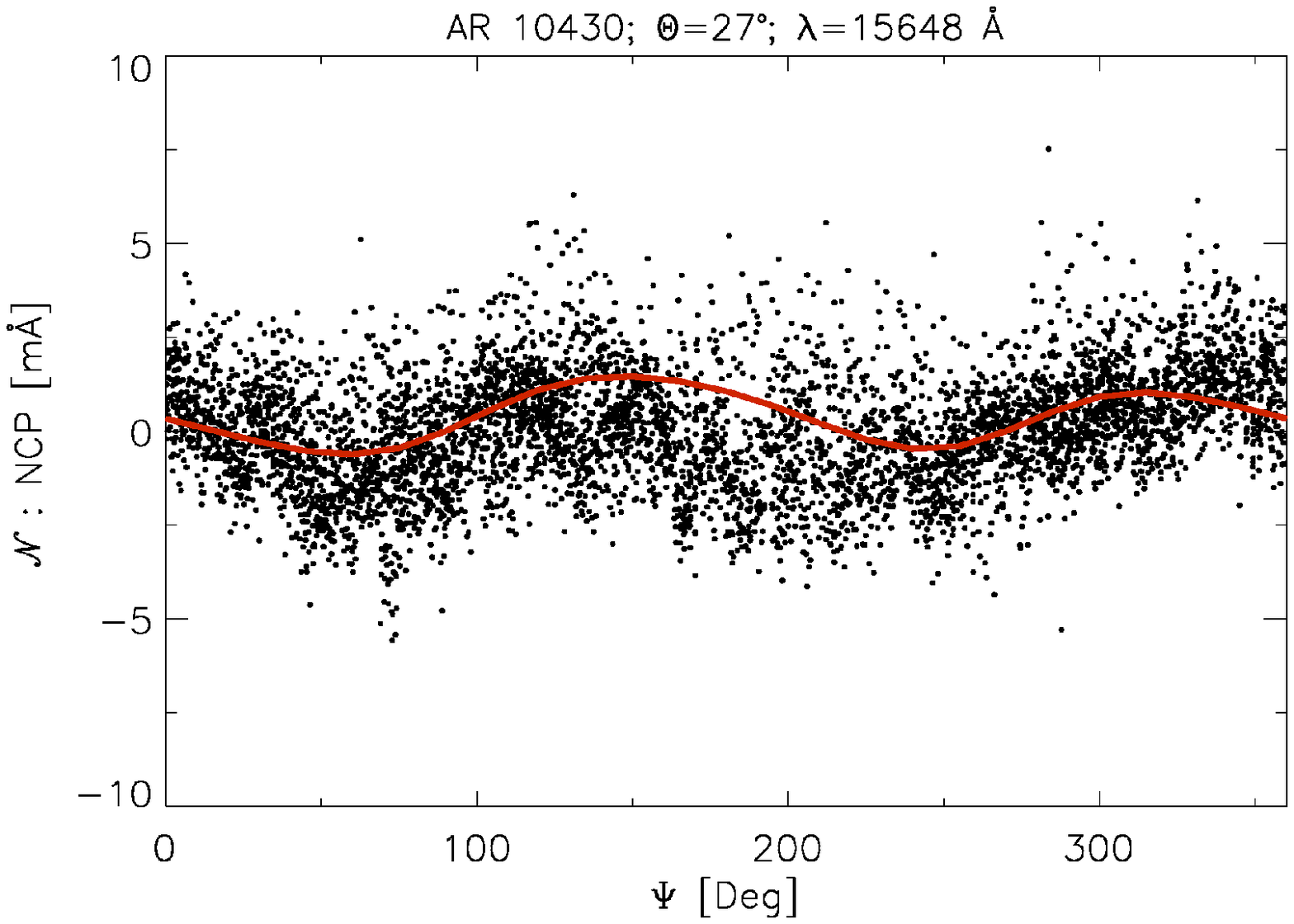}
\includegraphics[angle=0,width=7.5cm]{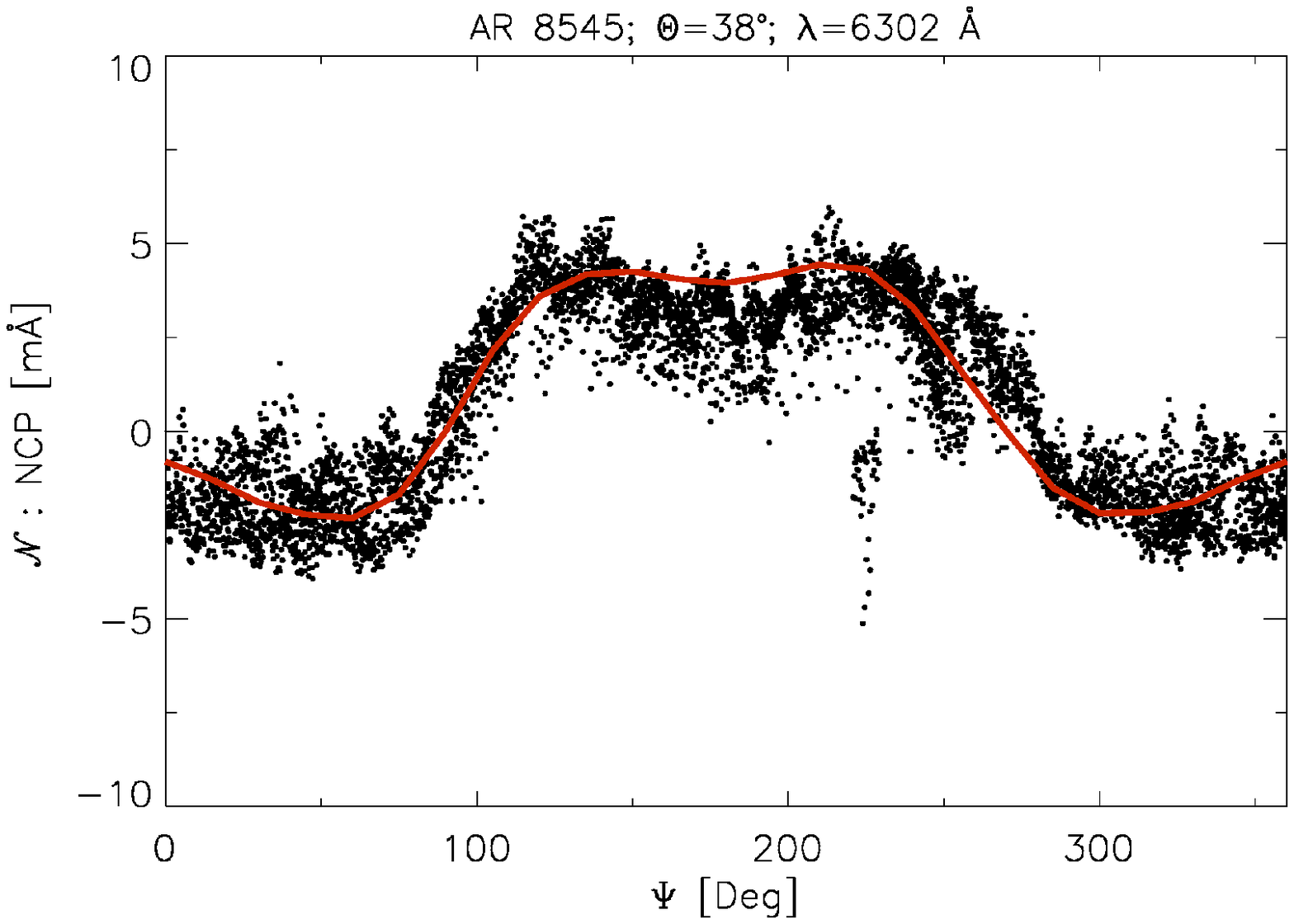}
\figcaption{Comparison between the observed (solid circles)
,and predicted (red solid lines) azimuthal variations of the
Net Circular Polarization, for two sunspots at heliocentric
angles: $\Theta=27^{\circ},38^{\circ}$.}
\end{center}

\begin{center}
\includegraphics[angle=0,width=7.5cm]{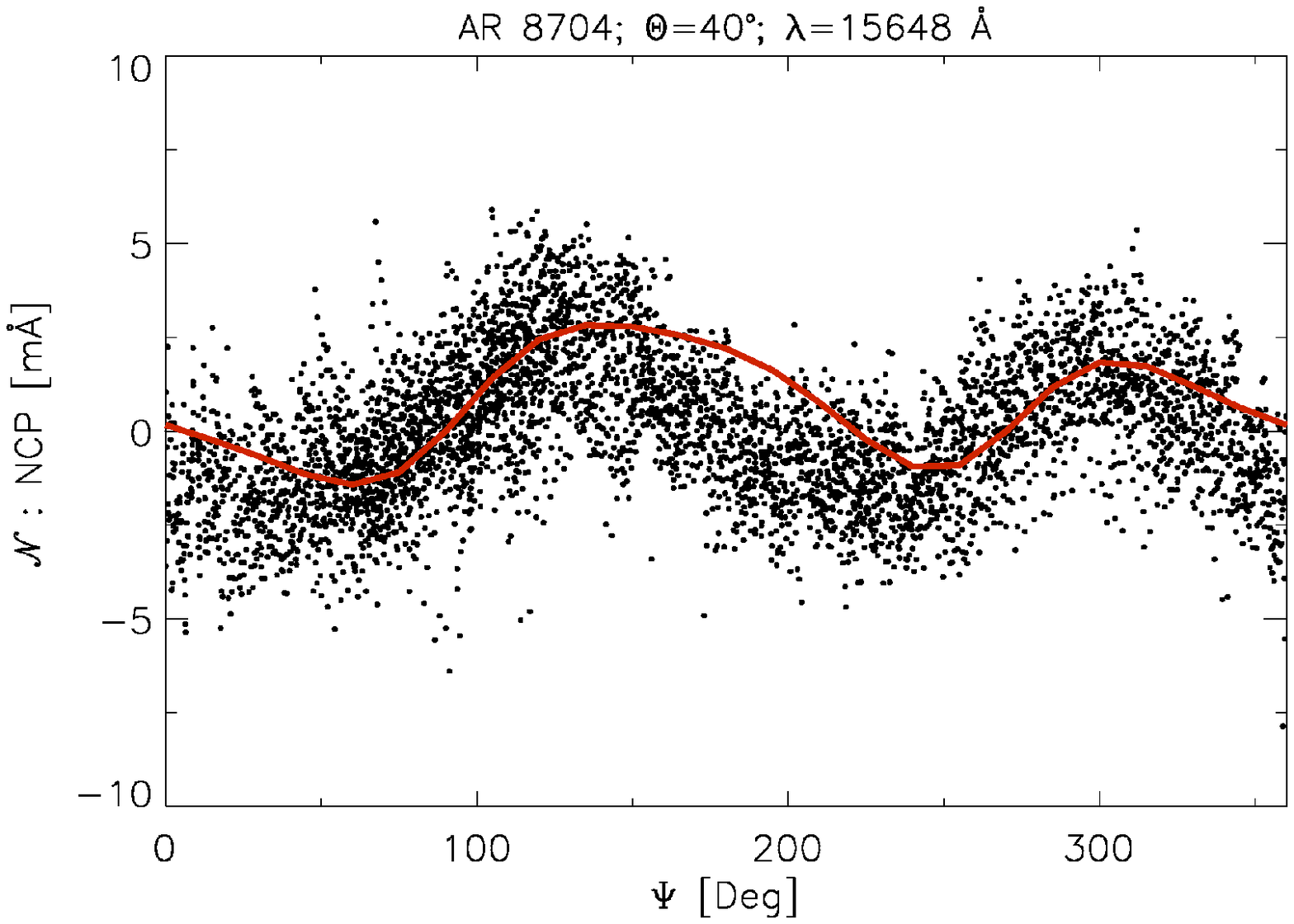}
\includegraphics[angle=0,width=7.5cm]{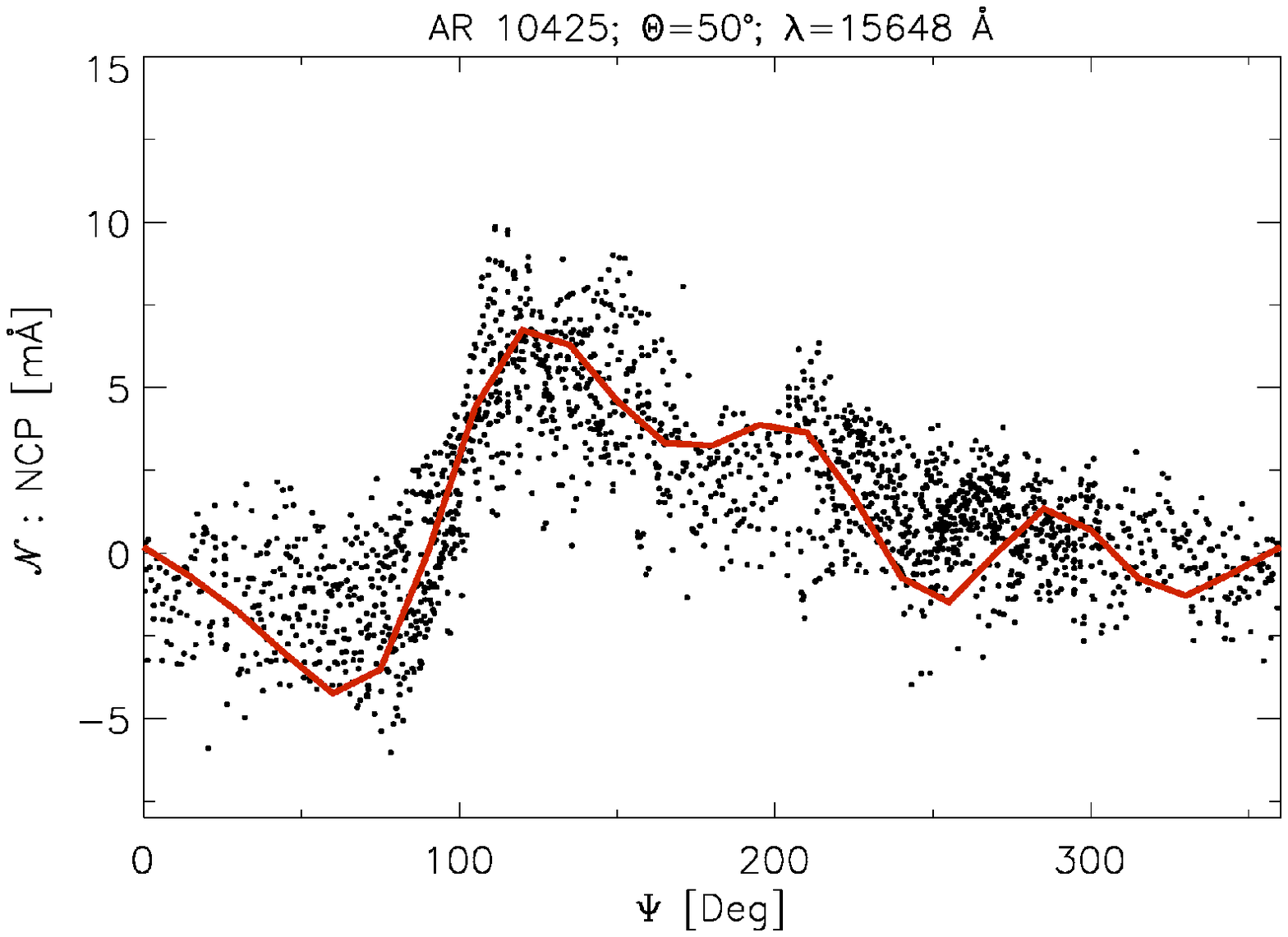}
\includegraphics[angle=0,width=7.5cm]{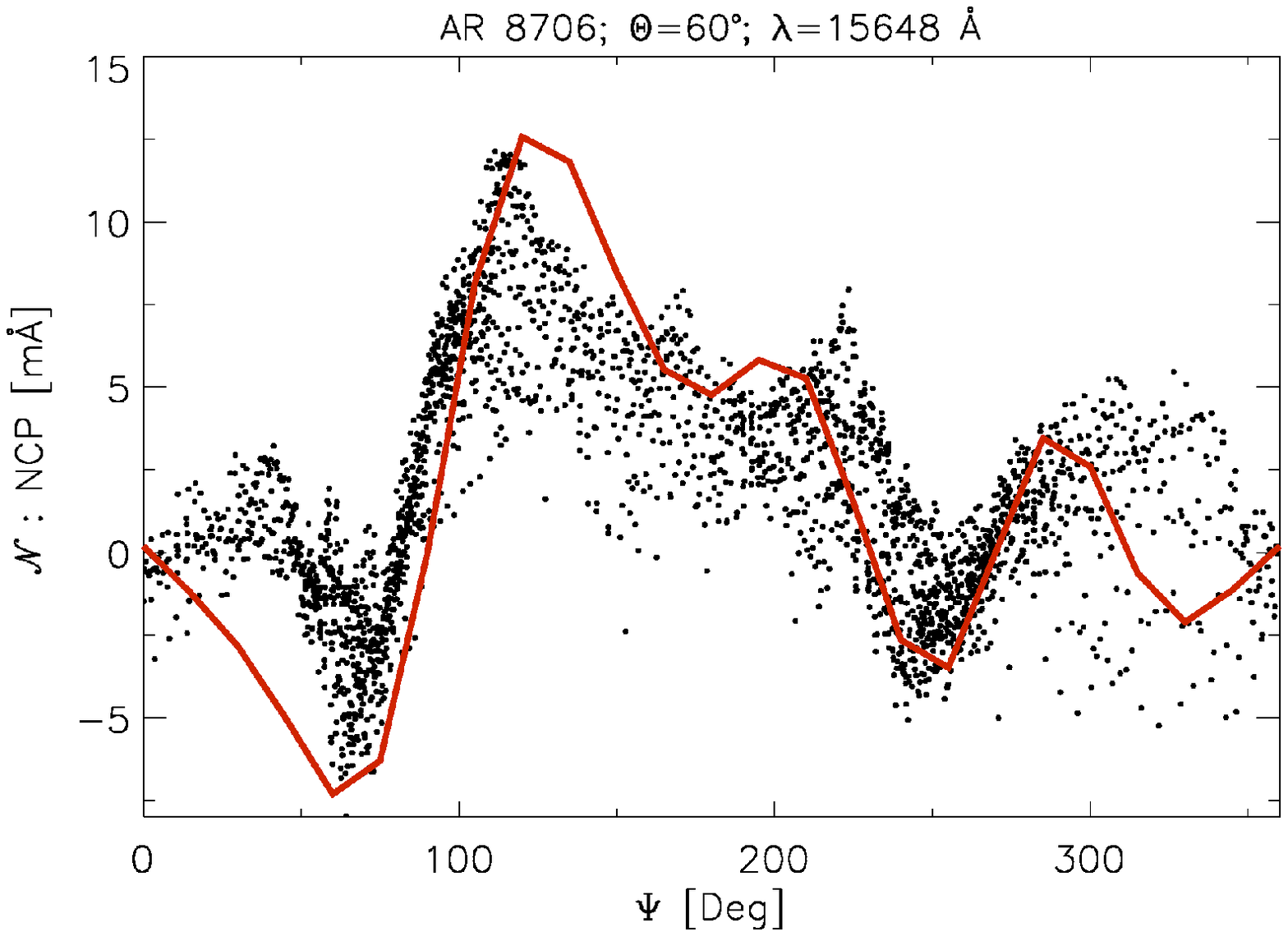} 
\figcaption{Same as Fig.~3 but for three
sunspots at $\Theta=40^{\circ},50^{\circ},60^{\circ}$.}
\end{center}

\section{Conclusions}

We have employed the model of Borrero (2007) to
predict the behavior of the Net Circular Polarization 
in the sunspot penumbra. This model finds the equilibrium 
configuration of a horizontal flux tube with circular cross 
section, that carries the Evershed flow,
and is embedded in a atmosphere with a potential magnetic field
pointing towards a different direction. Energy transfer
is neglected and therefore we cannot address how the penumbra 
is heated. This model consistently considers how the external 
magnetic field opens and bends in order to accommodate the horizontal flux 
tube. This generalizes the work of Solanki \& 
Montavon (1993), Mart{\'\i}nez Pillet (2000), M\"uller et al. 
(2002; 2006) and Schlichenmaier et al. (2002), by considering 
the 3D geometry of the problem.

We have compared our predictions with the observed NCP
in two neutral iron lines and in five different sunspots.
The agreement between theory and observations is remarkable, 
improving previous determinations based on simpler realizations 
of the uncombed penumbral model. Other models for the penumbral
fine structure (S\'anchez Almeida 2005; Scharmer \& Spruit 2006) should 
also try to explain these observations. In the future, we will attempt 
to reproduce also the full polarization profiles of these 
spectral lines (cf. Borrero et al. 2005,2006; Bellot 
Rubio et al. 2004).

\acknowledgements{We wish to thank Daniel Cabrera Solana for kindly providing 
some of the observations used in this work.}

\end{document}